# Multi-photon Absorption in Optical Pumping of Rubidium

Xinyi Xu (ID PIN:A51481739)
Department of Physics and Astronomy
Michigan State University

**Abstract**: In optical pumping of rubidium, a new kind of absorption occurs with a higher amplitude of radio frequency current. From measurement of the corresponding magnetic field value where this absorption occurs, there is a conclusion that it is multi-photon absorption. Both the degeneracy and energy of photons contribute to the intensity.

## I. INTRODUCTION

According to quantum mechanics, in the presence of a weak magnetic field, the amount of energy level splitting, or Zeeman Effect, is proportional to the value of the external magnetic field. The energy difference between sublevel M and M+1 can be calculated as

$$W(M+1) - W(M) = \Delta W = \mu_0 g_F B \quad (1)$$

where W is the interaction energy, $\Delta W$ is the energy splitting, $\mu_0$ is magnetic constant (vacuum permeability), $g_F$ is the g factor considering the interaction with the nucleus.

Optical pumping is a method to explore the atomic structure of atoms. An external magnetic field causes Zeeman Effect, and creates sublevels based on the Hyperfine Splitting. The atoms used in this experiment are rubidium atoms. Electronic configuration in the standard notation is

$$1s^2 2s^2 2p^6 3s^2 3d^{10} 4s^2 4p^6 5s.$$

With circularly polarized resonance radiation, a rubidium atom can be excited and deexcited between $^2S_{1/2}$ ground state and the $^2P_{1/2}$ excited state.

Take what occurs to atoms at F=1, M=0 sublevel of $^2S_{1/2}$ ground state as an example of optical pumping. By absorbing a photon, ground-state atoms in the F=1, M=0 sublevel jump to the F=1, M=1 sublevel of the excited state; atoms in the excited state decay to either M=0 or M=1 ground-state sublevels. Eventually, all atoms will be at F=1, M=1 sublevel of ground state and stop absorption [2]. If an atom at ground-state absorb an external photon and then deexcites to the F=1, M=1 sublevel from the F=1, M=0 sublevel, the thermodynamic equilibrium is broken, and this atom will absorb incoming light again. That is the whole process of optical pumping.

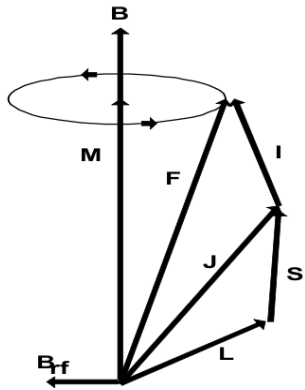

Figure 1. Vector relationship of magnetic fields and angular momenta components involved in the experiment, [1] where B is the static applied magnetic field, $B_{rf}$ is the varying radio frequency (RF) magnetic field, F is the total angular momentum of the atom, M is the component of F along the applied magnetic field, I is the nuclear spin angular momentum. S is the electron spin angular momentum. L is the electron orbital angular momentum.

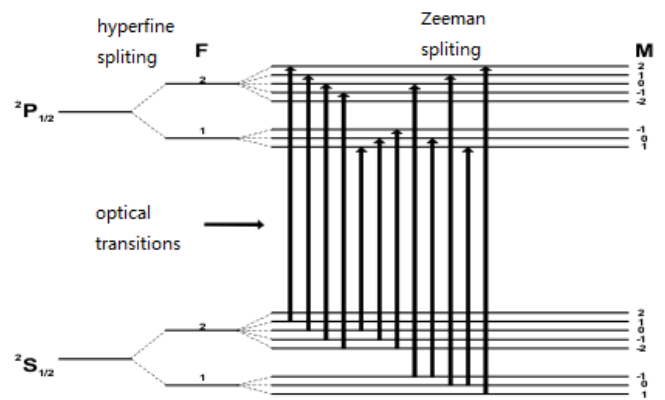

Figure 2. The transitions involved in optical pumping of Rb87, and the hyperfine splitting and Zeeman splitting we used.

The rubidium atoms absorb photons and get energy of ($\mu_0 g_F B$) to break thermodynamic equilibrium.

Traditionally, optical pumping experiments focus on single photon absorption, whereas we analyze multi-photon absorption. Take two-photon absorption (TPA) as an example of multi-photon absorption. The absorption of two photons with energy of $h\upsilon$ couldn't occur simultaneously. After absorbing the first photon, the atom reach a virtual energy state. This state is not an eigenstate and unstable. The atom would either continue to absorb a second photon to another eigenstate or emit a photon to the original eigenstate. [3] The probability of TPA should be much smaller than the probability of single-photon absorption. Only if an atom is located among enough photons can this double absorption, or double collision, occurs.

In multi-photon absorption, the relative energy levels of the ground electronic state and the resonance transition frequency can be calculated as

$$N h\upsilon = \mu_0 g_F B \qquad (2)$$

where $\upsilon$ is the transition frequency in Hz, h is Planck's constant, and N=2,3,4…

When rubidium absorption cell is considered as a medium, for single-photon absorption,

$$I_{out}=I_{in} \exp(-n\sigma L) \quad (3)$$

where $I_{out}$ is the intensity of transmitted light, $I_{in}$ is the intensity of incoming light, n is the volume density (concentration) of the medium, σ is the absorption cross-section and L is the thickness of medium. For double-photon absorption, the intensity after the sample is

$$I_{out} = I_{in} \frac{(1-R)^2 e^{-\alpha l}}{1 + \frac{\beta}{\alpha} I_{in}(1-R)(1-e^{-\alpha l})} \quad (4)$$

where α= nσ, β is the coefficient responsible for the two photon absorption. [4][5] RF amplitude influence n and σ, then influence $I_{out}$.

## II. METHODS AND THE SETUP

In laboratory, a radio frequency signal (RF) was used to excite coils around the rubidium absorption cell to produce the $B_{rf}$ of Figure 1. A specific frequency and amplitude of $B_{rf}$ was used. The current in an additional set of coils was varied to modify B of Figure 1, the static applied magnetic field, or horizontal magnetic field. A beam of light (photons) at a certain wavelength of 780nm passed through a vapor of the Rb atoms, and the intensity of transmitted light is measured. With $B_{rf}$=0, the rubidium atoms reached thermodynamic equilibrium. Such condition that no incoming light was absorbed providing a transmitted photon signal that is maximal and stable. When the whole magnetic field value B is the corresponding value of RF frequency υ, i.e.

$$B = N h\upsilon / (\mu_0 g_F) \quad (5)$$

atoms were pumped to another ground state sublevel and could be excited by incoming light. The transmitted intensity decreased because the absorption increased. Thus, a sharp drop of signal can be observed.

A rubidium discharge lamp were used to create the necessary light (photons). Including an interference filter (diameter 50mm), a linear polarizers in rotatable mounts (diameter 50mm) and a quarter wavelength plate in rotatable mount to get polarized light of 780nm wavelength. The polarized light was passed through a cell with rubidium vapor (Rubidium Absorption Cell in Figure 3). The photons passing through the rubidium vapor were detected using a silicon photodiode from Photonic Detectors Inc. PDB-C108. The light emitted by deexicitation of the excited states is isotropic and too weak to influence signal, so we assume detector only detects light from Rb discharge lamp. Two plano-convex lenses (diameter 50mm, focal length 50mm) are also used to minimize spherical aberrations. The apparatus arrangement on optical axis is shown as Figure 3. They should be adjusted at the same height before the experiment started.

Outside absorption cell are three pairs of Helmholtz coils (see Fig. 4), consisting of the Vertical Field Coils, the Horizontal Field Coils, and the Horizontal Sweep Field Coils (Sweep Field Coils). For all coils, the magnetic field value is proportional to current. During measurements, the hardware Figure 4(a) is covered by a black box to minimize stray signals from ambient light.

Other main equipment in this experiment includes FLUKE 6060B Synthesized RF Signal Generator (used to create radio frequency signal), TDS 3000B Series Digital

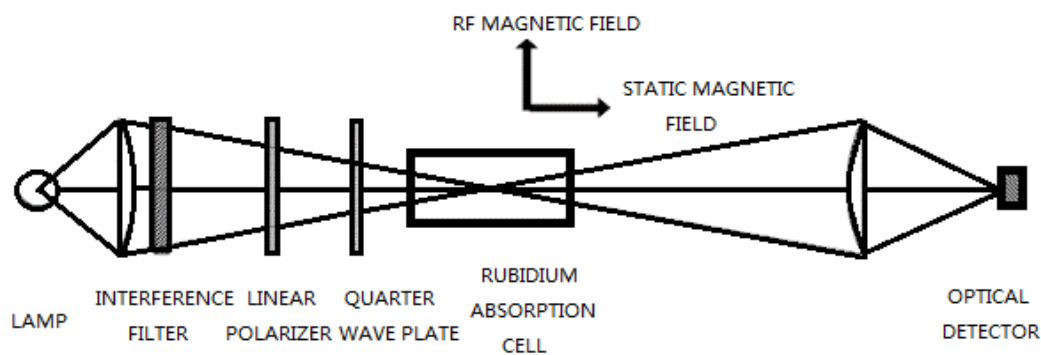

Figure 3. The apparatus arrangement for optical pumping. The optics on optical axis

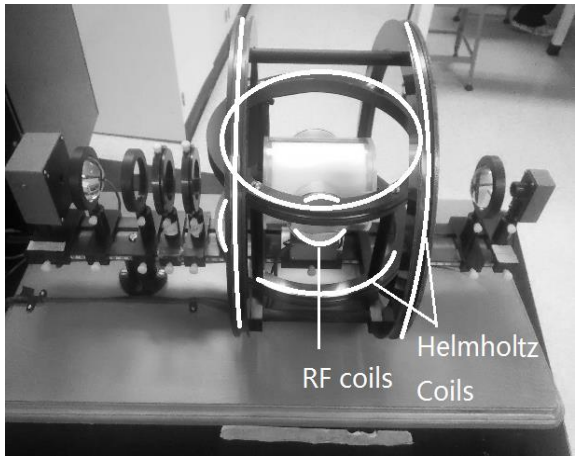 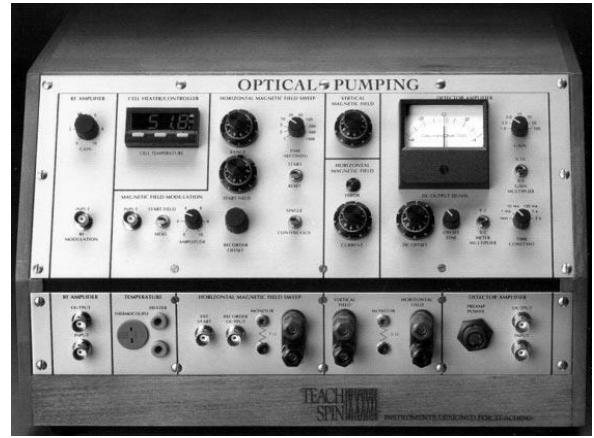

Figure 4. The apparatus arrangement for optical pumping. (a) Left: Helmholtz coils and RF coils around rubidium absorption cell. (b) Right: The front panel of main equipment. (http://teachspin.com /instruments /optical_pumping/index.shtml)

Phosphor Oscilloscopes, Teachspin Optical Pumping Apparatus, and three digital voltmeters.

The amplitude of RF in this experiment was shown on FLUKE 6060B Synthesized RF Signal Generator. Amplitude is described with unit dBm. The maximum it can reach is 19dBm, which was used frequently. Amplitude of more than 13dBm was uncalculated. The relationship between actual amplitude of RF coils' current (x) and digital amplitude shown on the signal generator (y) was experimentally determined to be

$$y = 9.9105\ln(x) - 42.458, \qquad (6)$$

Nevertheless, because the actual amplitude of current was found to change randomly when the digital amplitude was increased from 18.0dBm to 19.0dBm, and because there were systematic errors, the real amplitude of RF coils' current was used.

The rubidium vapor cell temperature was set at 50.0℃. The effect of the external magnetic field was minimized by adjusting both the direction of optical axis, and the Helmholtz Vertical Field.

## III. MEASUREMENTS AND RESULTS

## A. Optical Pumping and Single-photon Absorption at Lower Radio Frequency /Lower Magnetic Field

RF frequency was limited to be $<10^6$Hz (1MHz). At a specific RF frequency, the light intensity was measured as a function of the primary $B_{rf}$ field (see Figure 1). Sharp reduction in light intensity signaled photon absorption at several points (see Figure 5). When the sum of magnetic field caused by current in both main coils and sweep coils satisfy

$$B = \mu_0 g_F / (h\upsilon) \qquad (7)$$

Then a rubidium atom can be excited by a photon with energy ($h\upsilon$).

Single-photon absorption at lower radio frequency/magnetic field is a traditional optical pumping experiment. It can be used for calibration.

## B. The Magnetic Field of All Dips and Multi-Photon Absorption

When the amplitude of RF is increased sufficiently, some new dips appear. Figure 6

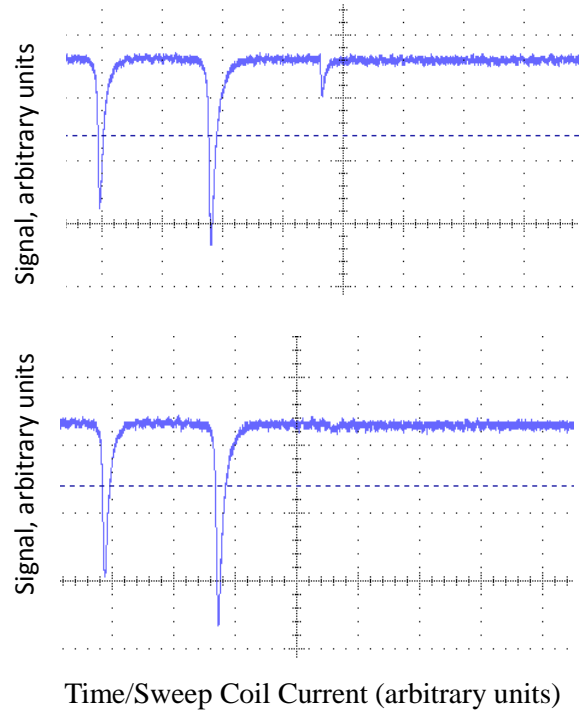

Time/Sweep Coil Current (arbitrary units)

Figure 6. Signals when $B_{rf}$ frequency was 0.3MHz. Upper graph is for the case of an RF amplitude of 19dBm; the lower graph was measured with RF amplitude of 18dBm.

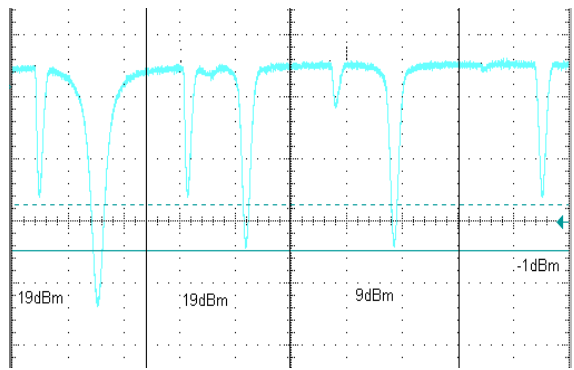

Figure 5. Frequency=0.4MHz, the dips of $^{87}$Rb with decreasing RF amplitude on oscilloscope. Right ones in every region are caused by unavoidable sudden decrease of sweep current, and they are not related to this experiment.

shows one of these new dips. As we can see, the small dip almost disappear when amplitude if 18dBm. The dips had been observed in the former experiments will be called Regular Dips (RDs) in the following; these dips appeared only with high RF amplitude will be called Amplitude Dips (ADs).

Higher amplitude equals to higher total energy. When one photon's energy remain $h\upsilon$ with fixed frequency, there should be more photons inside the rubidium vapor cell. The mean free path was shortened and this progress, eventually, enlarge the probability of collision between photons and rubidium atoms increased, and an atom even had an opportunity to absorb more photons. Hence, most possible hypothesis is that ADs are caused by multi-photon absorption.

If multi-photon absorption is what occurs in the rubidium cell, the corresponding magnetic field value must satisfy the equation (5)

$$B = N\,h\upsilon/(\mu_0 g_F).$$

or B should be proportional to RF frequency, and the slope should be $[Nh/(\mu_0 g_F)]$. Theoretical $g_F$ values of $Rb^{85}$ and $Rb^{87}$ are 1/3 and 1/2. When N=2,

$$2h\upsilon = \mu_0 g_F\, B \qquad (8)$$

$$B = 2h\upsilon/(\mu_0 g_F) \qquad (9)$$

Using $\mu_0/h = 1.3996$ MHz/gauss and $g_F = 1/2$, theoretical slope should be

$$2\,h/(\mu_0 g_F) = 2.8580\,(\text{Gauss/MHz}).$$

Figure 7 shows the measured relationship between magnetic field value and resonance RF frequency of the first AD. It gave a slope of 2.859 Guass/MHz. There is a difference of 10e-3Gauss/MHz or about 0.036% between calculated slope (2.8580Guass/MHz) and measured slope. Therefore, there is a strong possibility that the first "amplitude dip" was caused by double-photon absorption of $Rb^{87}$ atoms in weak external magnetic field, and the differences were caused by systematic errors.

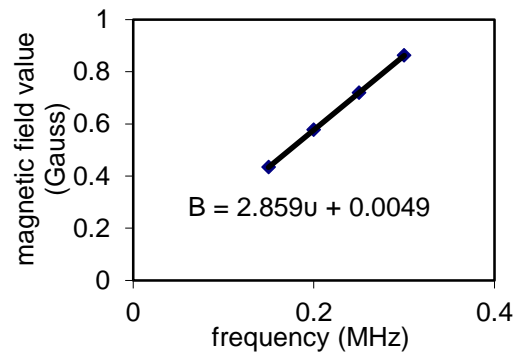

Figure 7. Magnetic field and resonance frequency of the first amplitude dip.

| Dip | Type | slope | intercept | R^2 | N | Isotope/$g_F$ | Theoretical Slope | Error (%) |
|---|---|---|---|---|---|---|---|---|
| 1 | RD | 1.4056 | 0.0090 | 0.9997 | 1 | Rb87 | 1.4290 | -1.63751 |
| 2 | RD | 2.1379 | 0.0036 | 0.9999 | 1 | Rb85 | 2.1437 | -0.27056 |
| 3 | AD | 2.8420 | 0.0057 | 0.9999 | 2 | Rb87 | 2.8580 | -0.55983 |
| 4 | AD | 4.2485 | 0.0184 | 0.9999 | 2/3 | Rb85/Rb87 | 4.2870 | -0.89806 |
| 5 | AD | 5.7408 | -0.0041 | 0.9999 | 4 | Rb87 | 5.7160 | 0.43387 |
| 6 | AD | 6.4539 | -0.0039 | 1.0000 | 3 | Rb85 | 6.4304 | 0.365452 |

Table 1. Linear fitting results of all the six dips above. Equation (3) was used to calculate theoretical value.

Then, we checked whether the following ADs could be explained by multi-photon absorption. Two RDs were measured as a reference. Figure 8 shows the measured relationship between magnetic field value and resonance RF frequency of the first six dips. Using equation (5), magnetic field values of all the dips are calculated. Table 1 shows the results and differences. From the comparisons, we can also know the reason of every AD, including the number of photons absorbed and isotope.

According to the last column of Table 1, the errors is less than 1%; the randomly distribution of errors means they are systematic errors. Hence, we proved the hypothesis that regular dips are caused by

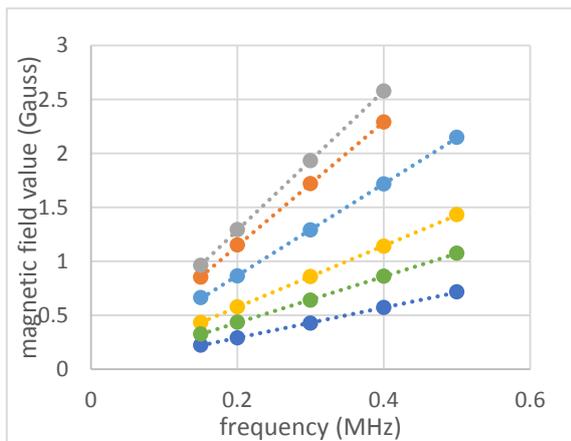

Figure 8. From bottom to top are magnetic field values and resonance frequency of the first six dips (amplitude of RF was 19dBm).

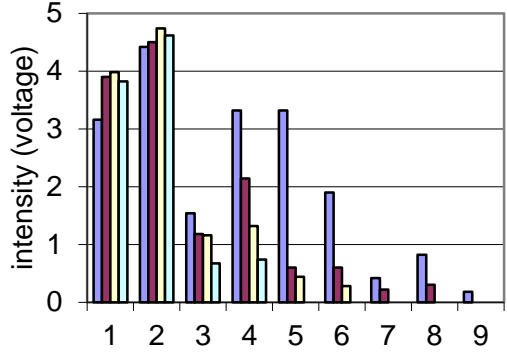

Figure 9. The intensity of dips (amplitude=19dBm). Horizontal axis shows the number of dips we found when magnetic field was increasing. For every dip, from left to right are intensity at 0.15MHz, 0.30MHz, 0.40MHz single-photon absorption and amplitude dips are caused by multi-photon absorption.

### C. Intensity of RDs and ADs

#### a) Degeneration of dips

Generally speaking, two isotopes - $Rb^{85}$ and $Rb^{87}$, contribute to intensity of their own dips. Only when

$$k=4j \qquad (10)$$

or

$$N \text{ for } Rb^{87}=3m \qquad (11.1)$$

$$N \text{ for } Rb^{85}=2n \qquad (11.2)$$

where k is the number of dips, including both RD and AD, and j, m, n=1,2,3… degeneracy occurs. The $(4j)^{th}$ dips consist of the contribution of both two isotopes, and, therefore, the intensity is unusual.

The ratio of intensity of the first two dips shows the ratio of percentage of $Rb^{87}$ and $Rb^{85}$. Theoretically, the ratio of intensity of two dips with the same N should be the same regardless of N. Assume $I_n$ is the intensity of the $n^{th}$ dip. Use the conclusion of Table 1, we should have been able to calculate $I_4$ with $I_1$, $I_2$, $I_3$ and $I_6$. However, whether linear relationship or quadratic relationship was used, the calculated results have differences of more than 20% with measured results. Supposing the results obey the Gaussian distribution, systematic errors are too large.

The hypothesis is that the relationship between percentage of isotope and intensity is complicated, because neither linear relationship nor quadratic relationship works when we calculate the intensity of degenerate dips. Whether it is the reality cannot be decided until experiments with more precise instrument be completed.

b) **How intensity change with amplitude of RF current**

From the AD in Figure 3 and the two RDs in Figure 4, it is obvious that the amplitude of RF current influence the intensity of both ADs and RDs. There is a hypothesis that the first is that $I_1$ and $I_2$ are proportional to amplitude. And double-photon absorption follows quadratic law, the intensity of three-photon absorption is proportional to the cube of amplitude, et cetera. That's to say,

$$I_N \propto A^N \qquad (12)$$

where N is the number of photons that a rubidium absorbed, $I_N$ is the intensity of a dip caused by N photons absorption and A is the amplitude of RF current.

From our measurement, for the 1st dip and 2nd dip,

$$I_1 = 3.6820\,A + 2.0235 \qquad (13)$$

$$I_2 = 3.6983\,A + 3.0831 \qquad (14)$$

and $R_1^2 = 0.9267$, and $R_2^2 = 0.9018$.

Nevertheless, the fittest relationships for the first two dips between actual amplitude (A) in unit mV and intensity ($I_n$) in unit mV are

$$I_1 = 0.9949 \ln(A) + 5.5108 \qquad (15)$$

$$I_2 = 0.9865 \ln(A) + 4.4589 \qquad (16)$$

$R_1^2 = 0.9941$, and $R_2^2 = 0.9789$. What's more, for amplitude dips, neither linear form nor exponent form seems fit. And polynomic form shows anything but a useful conclusion.

From the results above, relationship in equation (12) is disproved by our measurement. More experiments are needed before a new hypothesis is raised.

IV. **CONCLUSION**

The experiments described above are based on quantum mechanics and optical pumping of rubidium atoms. It gives a clear idea about Zeeman Effect and energy level of an alkali atom. It is useful when the energy difference is too large for single photon. Some problems related to intensity still need solving, including the intensity of degenerate dips and the relationship between intensity and amplitude.

## VI. CONTRIBUTIONS